\documentclass[sigconf]{acmart}

\setcopyright{none} 
\acmConference[Anonymous Submission to ACM CCS 2022]{ACM Conference on Computer and Communications Security}{Due 2 May 2022}{Los Angeles, CA}
\acmYear{2022}

\begin{document}
\title{Domain-Level Detection and Disruption of Disinformation} 

\author{Elliott Waissbluth, Hany Farid}
\affiliation{University of California, Berkeley}
\author{Vibhor Sehgal, Ankit Peshin, Sadia Afroz}
\affiliation{Avast Inc.}

\begin{abstract}
How, in 20 short years, did we go from the promise of the internet to democratize access to knowledge and make the world more understanding and enlightened, to the litany of daily horrors that is today's internet? We are awash in disinformation consisting of lies, conspiracies, and general nonsense, all with real-world implications ranging from horrific humans rights violations to threats to our democracy and global public health. Although the internet is vast, the peddlers of disinformation appear to be more localized. To this end, we describe a domain-level analysis for predicting if a domain is complicit in distributing or amplifying disinformation. This process analyzes the underlying domain content and the hyperlinking connectivity between domains to predict if a domain is peddling in disinformation. These basic insights extend to an analysis of disinformation on Telegram and Twitter. From these insights, we propose that search engines and social-media recommendation algorithms can systematically discover and demote the worst disinformation offenders, returning some trust and sanity to our online communities.
\end{abstract}

\ccsdesc[500]{Information systems~World Wide Web}

\keywords{disinformation; misinformation} 

\maketitle

\section{Introduction}

In the midst of the global pandemic, $20\%$ of the public believed Bill Gates was planning to use COVID-19 to implement a mandatory vaccine program with tracking microchips~\cite{nightingale2020examining}. Around the same time in mid-2020, the QAnon conspiracy ripped through social media, contending that a cabal of Satan-worshipping cannibalistic pedophiles and child sex-traffickers plotted against Donald Trump during his term as US President. A recent poll found $37\%$ of Americans are unsure whether the far-reaching QAnon is true or false, and $17\%$ believe it to be true~\cite{newall2020ipsos}.

The common thread in Bill Gates' COVID-microchips, QAnon's Satan-worshipping cannibals, and the long litany of conspiracies, lies, and general nonsense polluting the internet is the ease with which billions of online users can create and distribute content, the favoring by social-media's recommendation algorithms of the most outrageous and salacious content that drives user engagement, and the seemingly endless appetite of the general public for this content.

Today's disinformation campaigns are leading to real-world harms from vaccine hesitancy to the denial of the scientific consensus of the catastrophic effects of global climate change, denial of human-rights violations in Ukraine, and a lack of confidence in our electoral system.

Tackling disinformation on a per-post/image/video basis (e.g.,~\cite{li2018exposing,matern2019exploiting,li2020celeb,agarwal2019protecting,guera2018deepfake,shu2019role,shu2019beyond,ruchansky2017csi,tschiatschek2018fake,liu2018early,hanselowski2018retrospective,zhang2019detecting,girgis2018deep,shu2018understanding,reis2019supervised,aphiwongsophon2018detecting,karimi2018multi,shu2019detecting,aldwairi2018detecting,ma-etal-2017-detect,hounsel2020identifying,tang2021down}) is leading to a maddeningly massive game of online whack-a-mole. At the same time, recent studies have found that a relatively small number of users are responsible for the majority of online COVID~\cite{salam2022majority} and climate change~\cite{paul2021climate} disinformation, suggesting that tackling disinformation at a higher-level based on specific users, groups, or domains may be more effective.

By way of nomenclature, we refer to ``disinfo domains'' as those consisting of a broad category of domains that traffic in conspiracies, distortions, lies, disinformation, and more generally do not follow accepted journalistic standards, whether they are maintained by a state-sponsored actor, a private or public entity, or an individual. All other domains will be referred to as ``info domains.'' We describe in more detail in Section~\ref{subsec:domains} how domains are characterized as one or the other.

In this work, we describe a domain-level predictor of disinformation peddlers that classifies an entire domain (e.g.,~{\tt www.rt.com}) as an unreliable news source. This classifier relies on three signals: (1) hyperlinks (similar to~\cite{sehgal2021mutual}); (2) meta tags specified by the domain owners and used primarily for search-engine optimization; and (3) content in the form of the visible text on the landing and internally linked pages. Our primary contributions include:
\begin{enumerate}
    \item Analyzing a large set of $2500$ domains identified as trafficking in disinformation.
    \item Revealing a distinct and predictive patterns of hyperlinking, meta tag construction, and underlying content among disinformation peddlers in the form of accurate classifiers (F1=$94.1\%$) able to distinguish between disinfo and info domains.
    \item Building a hyperlink graph to reveal coordinated disinformation efforts.
    \item Extending our analysis to Telegram to discover the most prolific channels and users trafficking in disinformation, and using these identified channels to expand our disinfo domain data set.
    \item Extending our analysis to Twitter to discover the most prolific accounts trafficking in disinformation, and using these identified channels to expand our disinfo domain data set.
\end{enumerate}
%
%

\section{Related Work}
\label{sec:relatedwork}

According to a 2020 poll, slightly more than half of Americans rely on social media for at least some of their news~\cite{pew2020}. Given the growing flood of misinformation, lies, and conspiracies found on social media, recent efforts have focused on understanding the promotion and spread of misinformation on social media. Vosoughi et al.~\cite{vosoughi2018spread}, for example, analyzed the spread of misinformation on Twitter and found that misinformation spreads faster than the truth, and that misinformation is more novel than the truth and is designed to inspire a strong response of fear, disgust, and surprise, and hence more engagement in terms of likes, share, and retweets. Because of this, social media's recommendation algorithm tend to favor misinformation. Faddoul et al.~\cite{faddoul20} and Tang et al.~\cite{tang2021down}, for example, showed that YouTube's recommendation algorithm contribute to the spreading of conspiracies and misinformation. And, automated tools, such as Hoaxy~\cite{shao2018anatomy}, reveal in real time how misinformation spreads on Twitter. 

Related efforts have also analyzed the infrastructure support behind disinformation. Han et al.~\cite{han2021infrastructure}, for example, studied the service providers that power hundreds of disinformation and hate sites and found that disinformation sites disproportionately rely on several popular ad networks and payment processors, including RevContent and Google DoubleClick. In related work, by analyzing ads placed on over a thousand disinformation domains,~Zeng et al.~\cite{zeng2021polls} discovered that $42.5\%$ of the ads on these domains are political ads with a significant portion ($26.0\%$) being left-leaning.

Designed to inspire a strong response, the distinct linguistic characteristics of text-based misinformation have been used for automatic detection. A plethora of techniques have been proposed to automatically classify individual posts as misinformation or not. These range from leveraging classic machine learning techniques (SVM, LR, Decision Tree, Naive Bayes, k-NN) to more modern machine learning (CNN, LSTM, Bi-LSTM, C-LSTM, HAN, Conv-HAN) models to automatically detect misinformation (see~\cite{KHAN2021100032} for a comparative study of detection approaches). Having a large and accurately labeled list of misinformation, however, is difficult to obtain, which is why most of these studies use relatively small datasets.

Retroactive fact checking, however, is unlikely to stem the flow of misinformation. It has been shown, for example, that the effect of misinformation may persist even after false claims have been debunked~\cite{chan2017debunking,lewandowsky2017beyond}. In contrast, we focus our analysis at a higher level of determining if specific domains (e.g.,~{\tt www.rt.com}), Telegram channels, or Twitter users, are complicit in the spreading of misinformation. Detection at this level lends itself to earlier interventions in the form of demotion, demonetizing, and -- in the most extreme cases -- de-platforming.

We analyze the content of and the interconnection (through hyperlinking) of $2500$ disinfo domains -- identified by independent and nonpartisan sources -- to build a domain-level classifier of disinformation peddlers.

This type of hyperlink analysis has previously been examined. By analyzing $89$ news outlets, for example, Pak et al.~\cite{pak2020intermedia} found that partisan media outlets are more likely to link to nonpartisan media, but that liberal media link to liberal and neutral outlets whereas conservative media link more exclusively to conservative outlets. In analyzing hyperlinks between news media between 1999 to 2006, Weber et al.~\cite{weber2012newspapers} found that establishing hyperlinks with other, younger news outlets strengthens the position of that organization in the network thus boosting traffic. Hanley et al.~\cite{hanley2021no} observed a small-world phenomenon in QAnon domains in which only $22\%$ of authentic news websites hyperlink to a QAnon sites, as compared to $40\%$ of known disinformation domains.

While these previous studies were not focused broadly on disinformation, the recent work of~\cite{sehgal2021mutual} analyzed patterns of hyperlinking to classify domains as trustworthy or not. One crucial observation from this work is that disinformation domains are heavily connected to one another and loosely connected to authentic news domains. We expand on this work by analyzing more than twice as many domains, collated -- unlike in this earlier work -- from objective sources using standard metrics of journalistic standards. We also, unlike this earlier work, look at both the underlying HTML-based content as well as the hyperlinking connectivity. And, we expand our analysis of domain-level disinformation to Telegram and Twitter.

\section{Methods}

\subsection{Domains}
\label{subsec:domains}

We obtained two commercially available, manually annotated disinformation domain data sets. The first, from the Global Disinformation Index (GDI), is a monthly-updated list of global domains determined, using established journalistic standards, as being purveyors of disinformation. Our list, from Oct 2021, contains $1276$ disinfo domains. The second, from NewsGuard, is also a monthly-updated list of global domains rated and reviewed by trained journalists. Our list, from Oct 2021, contains $3403$ disinfo domains. Each site in this list is rated using nine basic, apolitical criteria of journalistic practice, from which each domain is scored on a scale of 0 (generally not trustworthy) to 100 (generally trustworthy). We labeled any site with NewsGuard's criteria that a score below $60$ is classified as a disinfo domain. The top $10,000$ Alexa-rated sites -- sans domains in our disinfo data set -- constitute our informational domains.  This info data set is, of course, a relatively crude comparison data set, but was selected because of their popularity and therefore overall online impact.

For each domain in our dataset, OpenWPM\footnote{\url{https://github.com/mozilla/OpenWPM}} was used to scrape the contents of the top-level domain (level 1), and to scrape internal pages linked by the top-level domain (level 2), etc., up to a maximum of $100$ pages. Any domain that returned a $404$ error were excluded from our analysis, yielding $2499/3689$ disinfo domains and $7888/10000$ info domains. These domains were further narrowed because of scraping blocks or primarily non-English content, yielding a final domain count of $2435$ disinfo domains and $5396$ info domains.

\subsection{Hyperlinks}
\label{subsec:hyperlinks}

The HTML hyperlink tag ({\tt <a href="..." </a>}) is used to link to an internal or external page. From the domains described above, we extracted all hyperlinks to external pages. An unweighted, directed graph of hyperlinks was then constructed in which the graph nodes are the top-level domains and a directed edge connects one domain that hyperlinked to another. The tldextract library\footnote{\url{https://github.com/john-kurkowski/tldextract}} was used to extract the top-level domain name from the URL. For example, if {\tt rt.com} hyperlinks to {\tt www.inforwars.com/posts/<...>}, then the graph contains a directed edge from {\tt rt.com} to {\tt infowars.com}.

Each domain is then featurized into a $3$-D vector consisting of the following ratios:
\begin{eqnarray*}
d_i & = & \frac{\mbox{\# incoming disinfo}}{\mbox{\# total incoming}} \\
d_o & = & \frac{\mbox{\# outgoing disinfo}}{\mbox{\# total outgoing}} \\
t   & = & \frac{\mbox{\# total incoming}}{\mbox{\# total outgoing}},
\end{eqnarray*}
where \# incoming disinfo is the sum of all unique disinfo domains linking to a specific domain (regardless of how many individual pages are linked),  \# outgoing disinfo is the sum of all unique disinfo domains a specific domain links to, and the total incoming count is only measured from our combined info and disinfo domains. Any feature for which the denominator is $0$ is reassigned a value of $-0.5$.

\subsection{Meta tags}
\label{subsec:metatags}

Meta tags are text-based HTML elements present in a web page's source code describing the page's content. These tags are mostly invisible to the user, as their purpose is to guide search engines. When a web architect creates a website, they take on the task of summarizing its content in meta tags to situate it among other relevant pages. Well crafted meta tags lead to better search engine optimization and user click through, motivating the architect to consider their meta tags carefully. Because of its importance in surfacing a web page, it is reasonable to assume that a peddler of disinformation will pay careful attention to this aspect of their page. 

From the the set of up to $100$ pages per domain, we scraped the content associated with seven meta tags: (1) {\tt keywords}, (2) {\tt description}, (3) {\tt og:title}, (4) {\tt og:keywords}, (5) {\tt og:description}, (6) {\tt twitter:description}, and (7) {\tt twitter:title}. These meta tags were selected because they tend to be descriptive of the underlying domain content, and because, unlike some other meta tags, were more likely to be specified in our domains ($61.6\%$ the domains in our dataset specified {\tt keywords}, $92.9\%$ specified {\tt description}, $86.1\%$ specified {\tt og:title}, $0.3\%$ specified {\tt og:keywords}, $84.3\%$ specified {\tt og:description}, $59.1\%$ specified {\tt twitter:description}, $60.3\%$ specified {\tt twitter:title}, and $96.9\%$ of domains specified at least one of these meta tags).

Each scraped domain yields a single string of concatenated words pre-processed to remove capitalization, punctuation, numbers, and stop words (e.g.,~{\em a, the, is, are}, etc.), and then stemmed using the Porter Stemmer algorithm~\cite{porter1980algorithm}, and lemmatized using the WordNet Lemmatizer~\cite{miller1998wordnet} (both of which are are implemented in the Python Natural Language Toolkit\footnote{\url{https://www.nltk.org}}). In order to reduce feature-vector dimensionality and focus on discriminatory features, any under-represented words (present in fewer than $10\%$ of our domains) or over-represented words (present in more than $90\%$ of our domains) are eliminated from consideration.

From the pre-processed meta-tag words, a unigram bag of words (BOW)~\cite{harris1954distributional} is weighted with each word's inverse document frequency~\cite{jones1972statistical} to yield a $12244$-D meta-tag feature vector for each domain. As described in Section~\ref{subsec:classification}, this feature vector is then pruned down to the top $500$ most diagnostic words.

\subsection{Content}
\label{subsec:content}

We define {\em content} as any visible text on a web page. This includes headers, footers, article titles, article content, link texts, button texts, etc. A website's visible text is likely to contain the majority of the linguistic information a user might engage with upon visiting a site.

From the the set of up to $100$ pages per domain, the visible text is extracted from the raw HTML by parsing out all text wrapped in any HTML tag, except for {\tt head}, {\tt meta}, {\tt title}, {\tt script}, and {\tt style}. The extracted content from all scraped pages within a domain is then concatenated into a single string and, as described in Section~\ref{subsec:metatags}, pre-processed to remove capitalization, punctuation, numbers, and stop words, then stemmed, and lemmatized. As before, any under-represented tokens or over-represented tokens are eliminated from consideration.

From the pre-processed words, a unigram bag of words (BOW) is weighted with each token's inverse document frequency to yield the final $80386$-D content-based feature of each domain. As described in Section~\ref{subsec:classification}, and as with the meta-tag features, this feature vector is then pruned down to the top $500$ most diagnostic words.

\subsection{Classification}
\label{subsec:classification}

Our classification objective is to distinguish between info and disinfo domains based on the domain-specific meta tag, content, and hyperlink features described in the previous sections, as well as from a combination of all three features. For the meta tag and content-based classifiers, we begin by reducing the original feature dimensionality of $12244$-D and $80386$-D to $500$-D. This is accomplished by training a logistic-regression classifier (using Python's scikit-learn library) and extracting the top $500$ most predictive distinguishing words from, separately, the original meta tag and content bag-of-words. 

For each feature, a separate linear support vector machine (SVM) is then trained (using Python's scikit-learn library) on a random $90\%$ of the info and disinfo domains, and tested on the remaining $10\%$ of the domains. A single combined classifier is also trained on all three features. Each component of the $500$-D meta-tag and content features are specified in the range $[0,1]$. Each component of the $3$-D hyperlink feature is normalized into the range $[0,1]$ (by subtracting the minimum possible value of $-0.5$ [see Section~\ref{subsec:hyperlinks}] and dividing by the maximum value of $10.5$). The SVM is optimized using a $5$-fold cross validation grid search across kernel functions, regularization parameter $C$, and $l_1$ and $l_2$ penalty functions, yielding a classifier with a linear kernel, an $l_2$ penalty function, and $C = 29.76$ for the $500$-D text-based features, and $C=10000$ for the $3$-D hyperlinking features.

\section{Results}

\begin{table}[t]
    \begin{center}
        \begin{tabular}{r|cccc}
            measure     & meta tags & content & hyperlinks & amalgamated \\
            \hline
            accuracy    & 94.4 & 95.7 & 95.1 & {\bf 96.3} \\
            precision   & 96.1 & 96.1 & {\bf 97.4} & 94.5 \\
            recall      & 85.5 & 89.8 & 86.6 & {\bf 93.7} \\
            F1          & 90.5 & 92.8 & 90.7 & {\bf 94.1} 
        \end{tabular}
    \end{center}
    \caption{Classifier performance (reported as percentages) for distinguishing info from disinfo domains, using individual features (meta tags, content, and hyperlinks) and these features combined into a single classifier (amalgamated). The bold-faced values correspond to the best performing classifier per measure.}
    \label{tab:results}
\end{table}
%
%
%
%

\subsection{Meta tags}

Shown in Table~\ref{tab:results} is the testing SVM (Section~\ref{subsec:classification}) performance averaged over $100$ random $90\%/10\%$ dataset training/testing splits using only the $500$-D representation of a domain's meta tags (Section~\ref{subsec:metatags}). With an average accuracy of $94.4\%$, and a minimum and maximum accuracy of $93.9\%$ and $94.9\%$, it would appear the meta tags are a reasonably good predictor of domain trustworthiness.

With an average (min/max) precision -- ratio of correctly classified disinfo domains to all classified disinfo domains -- of $96.1\%$ ($95.3\%$/$97.0\%$) and an average recall -- ratio of correctly classified disinfo domains to the total number of disinfo domains -- of $85.5\%$ ($84.0\%$/$87.0\%$), the classifier is somewhat biased. We hypothesize that this bias is due to more diversity in the info domains, as compared to the disinfo domains.

The top-ten most predictive (stemmed and lemmatized) words for classifying a domain as disinfo are:
\begin{quote}
    {\tt prank, archiv, conserv, gadget, ha, trump, freedom, polit, forb, russian}
\end{quote}
where {\tt archiv} is the prefix for such words as {\tt archive, archives, archiving}, {\tt ha} is a prefix for, among others, {\tt ha-ha}, {\tt polit} is the prefix for such words as {\tt politic, political, politician}, and {\tt forb} is a prefix for, among others, {\tt forbes}.

One benefit of the somewhat simplistic logistic regression classifier is it is interpretable allowing us to see that the classifier seems to be picking up on words that are intuitively related to disinformation.

\subsection{Content}

Shown in Table~\ref{tab:results} is the testing SVM (Section~\ref{subsec:classification}) performance averaged over $100$ random $90\%/10\%$ dataset training/testing splits using only the $500$-D representation of a domain's content (Section~\ref{subsec:content}). With an average accuracy of $95.7\%$, and a minimum and maximum accuracy of $95.3\%$ and $96.1\%$, the content is a slightly better predictor of domain trustworthiness than the meta tags. It is a little surprising that this classifier is only slightly more accurate than the classifier using only the meta tags ($95.7\%$ vs. $94.4\%$). We posit this is because the meta tags -- designed to drive search engine optimization -- consist of the most descriptive parts of a webpage's content. 

With an average (min/max) precision of $96.1\%$ ($95.3\%$/$96.9\%$) and an average recall  of $89.8\%$ ($88.7\%$/$90.9\%$), the classifier is slightly less biased than the meta tag classifier. 

The top-ten most predictive (stemmed and lemmatized) words for classifying a domain as disinfo are:
\begin{quote}
    {\tt prank, donat, petit, drudg, trunew, wordpress, immigr, post, freedom, would}
\end{quote}
where {\tt donat} is the prefix for such words as {\tt donation, donate, donating}, {\tt petit} is the prefix for such words as {\tt petition}, \linebreak {\tt petitioning}, {\tt petitioned}, {\tt immigr} is the prefix for such words as {\tt immigration}, {\tt immigrant}, {\tt immigrating}. The popular blogging site {\tt wordpress.com} often embeds the text ``Powered by WordPress'' in individual posts, and this domain also allows users to create profiles on other sites, tagging any comments on those sites with ``WordPress Profile.'' These are the likely sources of the predictive {\tt wordpress}.

\begin{figure}
    \centering
    \includegraphics[width=0.95\linewidth]{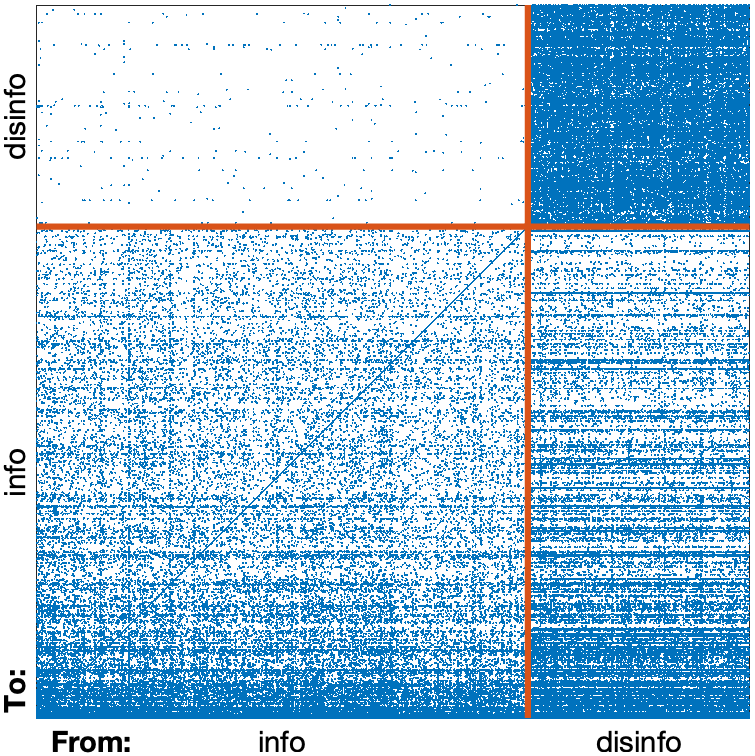}
    \caption{The hyperlinking adjacency matrix for the $7831$-node graph consisting of $5396$ info domains and $2435$ disinfo domains reveals a strong disinfo-disinfo (top right corner) and weak info-disinfo hyperlinking (top left corner).}
    \label{fig:adjmatrix}
\end{figure}

\subsection{Hyperlinks}

Shown in Figure~\ref{fig:adjmatrix} is the hyperlinking adjacency matrix for the $7831$-node graph consisting of $5396$ info domains and $2435$ disinfo domains (Section~\ref{subsec:domains}). There is a connection (blue dot) from one domain (horizontal axis) to another domain (vertical axis) if the from-domain contains at least one hyperlink to the to-domain. In this visualization, we can see that there are many hyperlinks from disinfo domains to other disinfo domains (upper right corner of the adjacency matrix). At the same time, info domains infrequently link to disinfo domains (upper left), and both info an disinfo domains link to info domains with similar frequency (lower left, and lower right corners of the adjacency matrix). The info domains on the horizontal axis are ranked -- from bottom to top -- in order of decreasing popularity (as determined by Alexa rankings), hence the gradient in hyperlinks with more links evident to the more popular domains.

\begin{figure*}
    \centering
    \begin{tabular}{c}
        \includegraphics[width=0.80\linewidth]{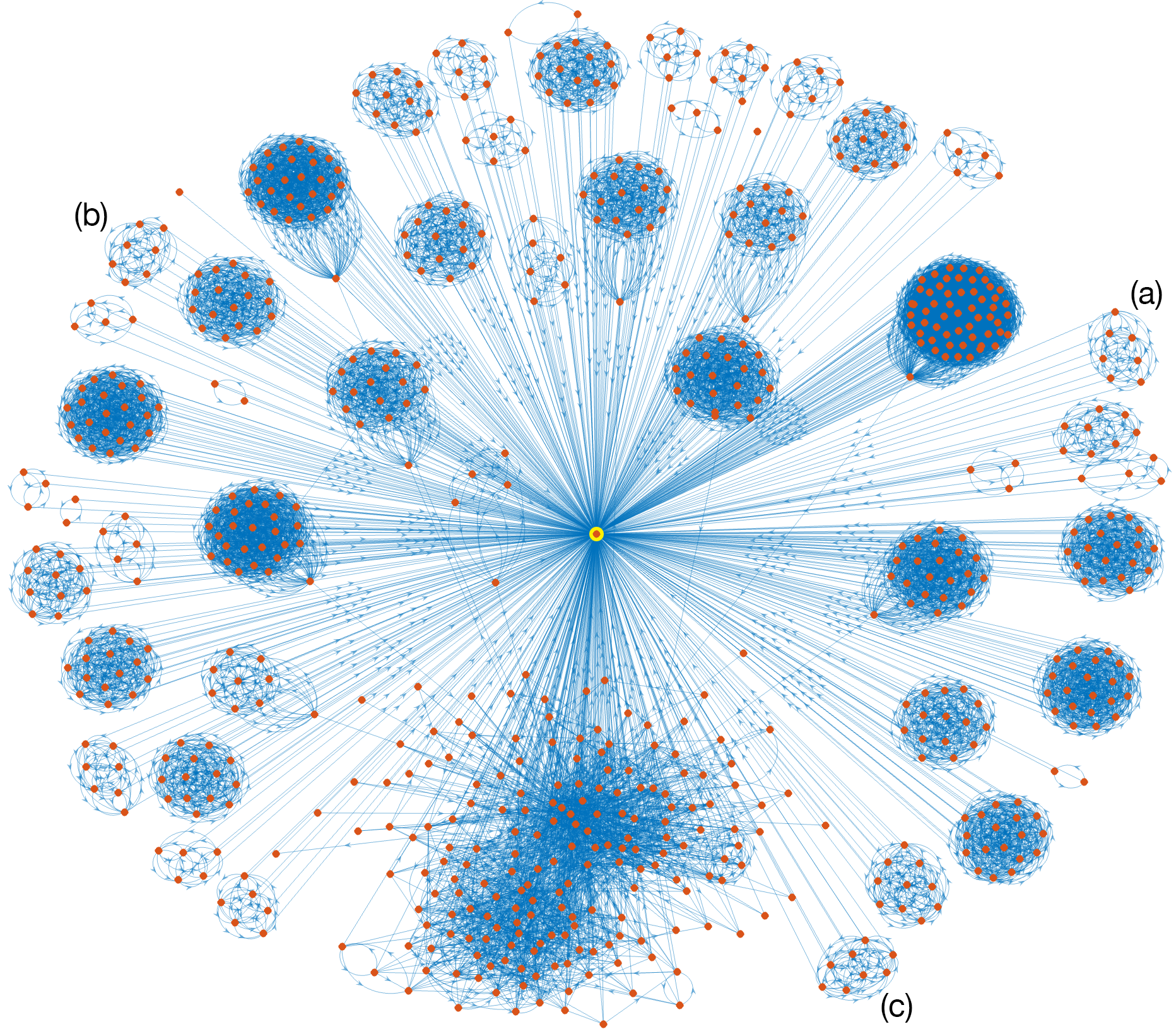}
    \end{tabular}
    \caption{Shown is the subgraph induced by the domain ({\tt www.freebeacon.com}) with the highest in-degree. The top-ranked domain is rendered with a yellow ring around the red node symbol. The small, fully-connected clusters correspond to imposter sites designed to look like local news sources (see Figure~\ref{fig:news-thumbnails} for a description of three of these clusters labeled (a)-(c)).}
    \label{fig:in_out_degree}
\end{figure*}

Shown in Table~\ref{tab:results} is the testing SVM (Section~\ref{subsec:classification}) performance averaged over $100$ random $90\%/10\%$ dataset training/testing splits using only the $3$-D representation of a domain's hyperlinking patterns (Section~\ref{subsec:hyperlinks}). With an average accuracy of $95.1\%$, and a minimum and maximum accuracy of $94.7\%$ and $95.5\%$, the hyperlinks are a comparable predictor of domain trustworthiness as compared to the meta tags ($94.4\%$) and content ($95.7\%$). The advantage, however, of this hyperlinking feature is that it is language agnostic and, as we will see below (Section~\ref{subsec:telegram} and~\ref{subsec:twitter}) seems to generalize to other services like Telegram and Twitter.

With an average (min/max) precision of $97.4\%$ ($96.8\%$/$98.0\%$) and an average recall of $86.6\%$ ($85.3\%$/$87.9\%$), the classifier, as with the meta tags and content, remains biased.

\subsection{Amalgamated}
\label{subsec:amalgamated}

Shown in Table~\ref{tab:results} is the testing SVM performance averaged over $100$ random $90\%/10\%$ dataset training/testing splits using the concatenated meta tag, content, and hyperlinking features as described in  Section~\ref{subsec:classification}.

With an average accuracy of $96.3\%$, and a minimum and maximum accuracy of $95.9\%$ and $96.7\%$, the combined classifier slightly outperforms the individual classifiers. With an average (min/max) precision -- ratio of correctly classified disinfo domains to all classified disinfo domains -- of $94.5\%$ ($93.5\%$/$95.4\%$) and an average recall -- ratio of correctly classified disinfo domains to the total number of disinfo domains -- of $93.7\%$ ($92.6\%$/$94.8\%$), the classifier is, unlike the individual classifiers, unbiased.

\begin{figure*}
    \centering
    \begin{tabular}{cc}
        \raisebox{1cm}{(a)} & \includegraphics[width=0.95\linewidth]{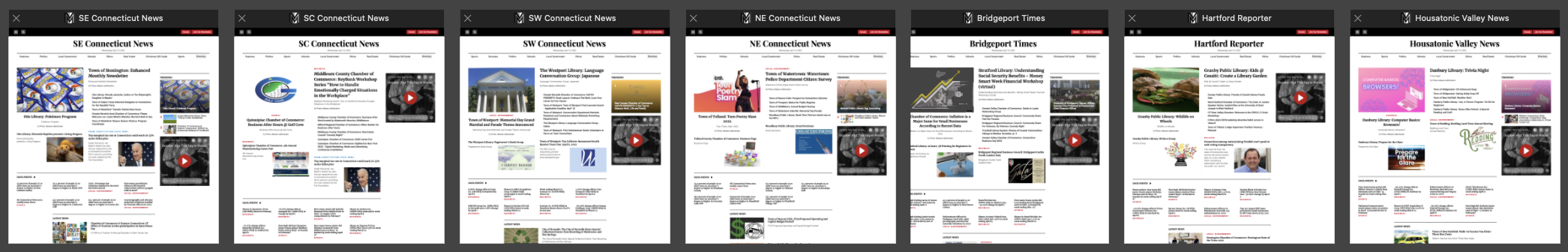} \\
        \raisebox{1cm}{(b)} & \includegraphics[width=0.95\linewidth]{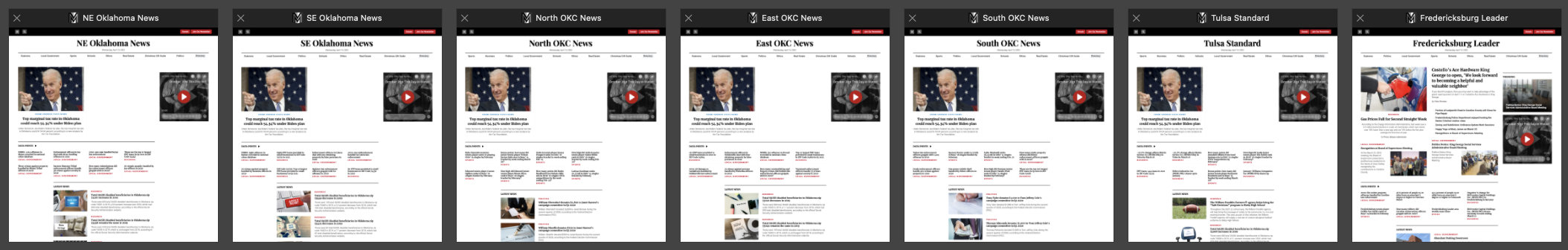} \\
        \raisebox{1cm}{(c)} & \includegraphics[width=0.95\linewidth]{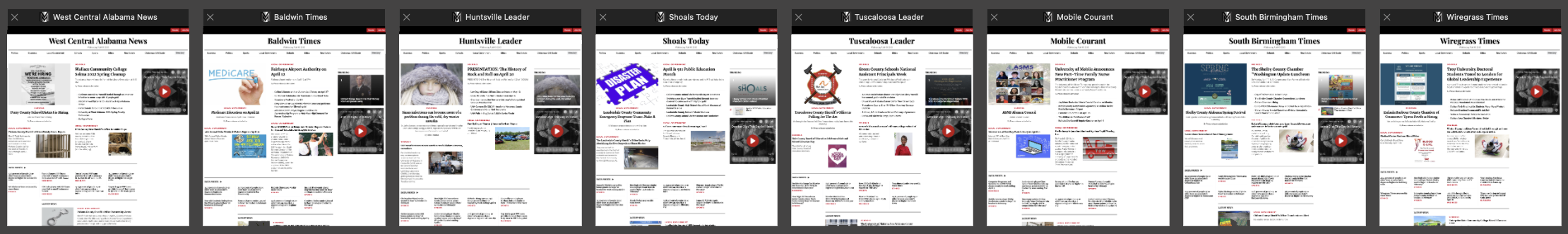} 
    \end{tabular}
    \caption{Thumbnails of three fully-connected subgraphs shown in Figure~\ref{fig:in_out_degree}. These imposter sites, all controlled by Metric Media News, are designed to look like local news sources.}
    \label{fig:news-thumbnails}
\end{figure*}
%
%

\section{Analysis and Insights}

\subsection{Networks}

The disinfo domains with the highest in-degree (i.e.,~many domains link to these) are:
\begin{enumerate}
    \item {\tt www.freebeacon.com}
    \item {\tt www.naturalnews.com}
    \item {\tt www.theepochtimes.com}
    \item {\tt www.zerohedge.com}
    \item {\tt www.brighteon.com}
\end{enumerate}
The disinfo doamins with the highest out-degree are (i.e.,~these domains have a large number of outgoing links):
\begin{enumerate}
    \item {\tt www.wakeupkiwi.com}
    \item {\tt www.conservapedia.com}
    \item {\tt www.moonofalabama.org}
    \item {\tt www.survivalinstitute.com}
    \item {\tt www.conservativerevival.com}
\end{enumerate}

Shown in Figure~\ref{fig:in_out_degree} is the subgraph of disinfo domains induced by the top-ranked in-degree {\tt www.freebeacon.com}. Launched in 2012, the Washington Free Beacon is an American political website ranked by Media Bias/Fact check (\url{mediabiasfactcheck.com}) as:
\begin{quote}
``Moderately to strongly biased toward conservative causes through story selection and/or political affiliation. They may utilize strong loaded words (wording that attempts to influence an audience by using appeal to emotion or stereotypes), publish misleading reports, and omit information that may damage conservative causes. Some sources in this category may be untrustworthy.  sources. Overall, we rate the Washington Free Beacon {\em Right Biased} based on story selection that favors the right and {\em Mixed} for factual reporting due to misleading and false claims.''
\end{quote}

A total of $735$ disinfo domains hyperlink to this top-ranked domain (only $35$ info domains, not shown in the subgraph, link to this top-ranked domain). This subgraph reveals a fascinating structure consisting of a large number of highly-connected sub-graphs, each of which hyperlinks to {\tt www.freebeacon.com} (shown in the center with a yellow ring around the red node symbol). The small compact clusters each correspond to purportedly regional news sites. The seven, fully-connected domains in Figure~\ref{fig:in_out_degree}(a), for example, consist of the Connecticut-based sites\footnote{{\tt www.seconnnews.com}, {\tt www.swconnnews.com}, {\tt www.scconnnews.com}, {\tt www.neconnnews.com}, {www.bridgeporttimes.com}, {\tt www.hartfordreporter.com}, and {\tt www.housatonicvalleynews.com}}. Similarly, the seven domains in the fully-connected Figure~\ref{fig:in_out_degree}(b) consist of Oklahoma-based sites\footnote{{\tt www.neoklahomanews.com}, {\tt www.seoklahomanews.com},  {\tt www.northokcnews.com}, {\tt www.eastokcnews.com}, {\tt www.southokcnews.com}, {\tt www.tulsastandard.com}, and {\tt www.fredericksburgleader.com}}, and the eight fully-connected domains in Figure~\ref{fig:in_out_degree}(c) correspond to Alabama-based sites\footnote{{\tt www.wcalabamanews.com}, {\tt www.baldwintimes.com}, {\tt www.huntsvilleleader.com}, {\tt www.shoalstoday.com}, {\tt www.tuscaloosaleader.com}, {\tt www.mobilecourant.com}, {\tt www.southbirminghamtimes.com}, and {\tt www.wiregrasstimes.com}}.

Shown in Figure~\ref{fig:news-thumbnails}(a)-(c) are thumbnails of these regional domains, revealing a similar layout, ads, and stories within and across the clusters. According to NewsGuard, all of these domains are controlled by Metric Media News which Media Bias/Fact check describes as:
\begin{quote}
``Overall, we rate Metric Media LLC right-center biased and Questionable based on a lack of transparency, the publication of false information, and nondisclosure of over 1000 imposter websites that are designed to look like local news sources.''
\end{quote}

Except for the larger cluster at the bottom of the graph in Figure~\ref{fig:in_out_degree}, each of the other tightly-coupled clusters belong to a local news sources controlled by Metric Media News.

Our analysis is capable not only of predicting if a domain is a likely peddler of disinformation, but also to reveal these types of sub-structures in the disinfo ecosystem. Of particular interest are these types of highly linked-to domains, but also the fully connected cliques revealing potentially coordinated disinformation efforts.

\begin{figure*}[t]
    \centering
    \includegraphics[width=0.75\linewidth]{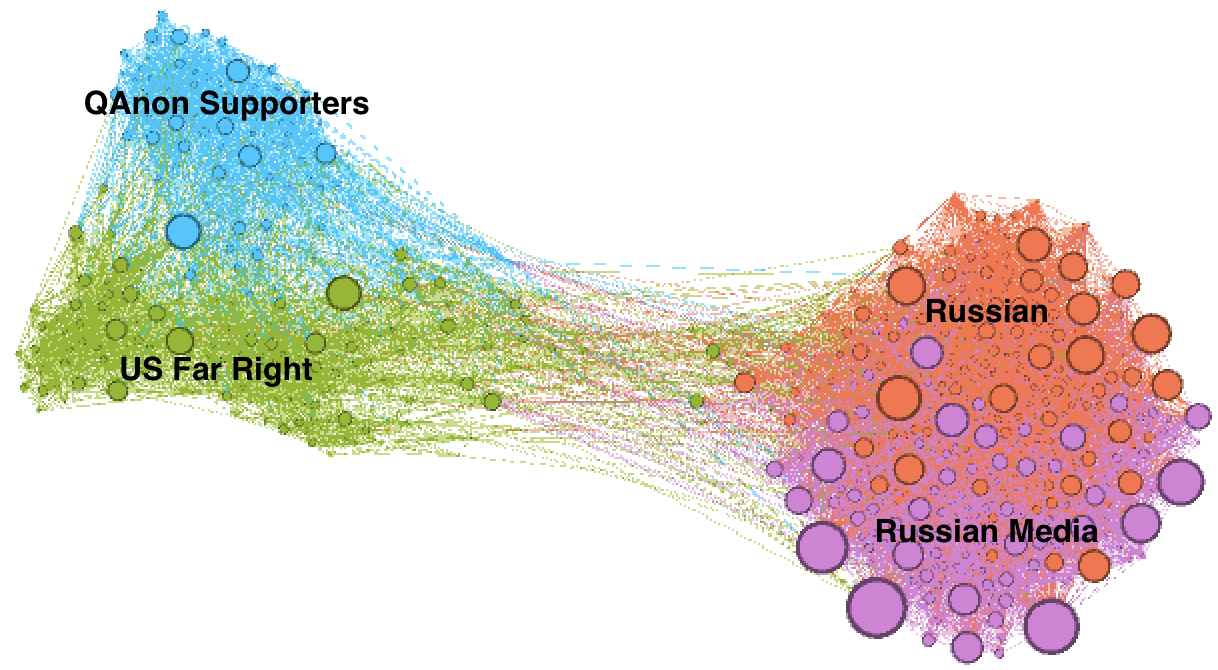}
    \caption{A subgraph of $860$ Telegram channels extracted from the full Telegram graph of $11,127$ channels constructed by considering only the channels with degree greater than $20$. An edge is added between nodes {\em A} and {\em B} if {\em A} forwarded a message from {\em B}. The size of the nodes represent the in-degree of the nodes (number of times the channel is mentioned by other channels). The nodes are color-coded according to their general category, revealing a striking relationship between Russian-based sites, QAnon, and US-based far-right groups.}
    \label{fig:telegram-communities}
\end{figure*}

Because Metric Media's imposter sites are so numerous, accounting for $36.1\%$ of our disinfo domains, we worried that their dominance skewed our classifier (Table~\ref{tab:results}). We, therefore, retrained our classifier using only one domain from Metric Media's vast network, and only one domain from any other identified (by NewsGuard or GDI) similar type of network. This yielded a total of $1061$ disinfo domains and a classification accuracy of $94.3\%$ (as compared to the previous $96.3\%$ accuracy), but a significantly reduced F1 score of $81.0\%$ (as compared to $94.5\%$). Although the precision remained relatively high at $89.1\%$ (as compared to $93.7\%$), the recall reduced significantly to $74.4\%$ (as compared to $94.1\%$). We hypothesize that this reduction in overall performance is due to a significantly smaller and more diverse dataset. We expect, however, that as we grow and diversify our disinfo data set, classification will further improve.

\subsection{Telegram}
\label{subsec:telegram}

Disinformation, of course, travels in many different forms and circles, from imposter news sites to social-media posts, public and private groups, and individual text messages. We wondered if our domain-level hyperlink analysis would generalize to some of these other mediums. We next describe a hyperlink analysis of disinformation networks on public Telegram channels.

The telegram channel {\tt Donbass Insider} has previously been implicated in advancing pro-Russian propaganda\footnote{The perceived threat of this and related channels led to Telegram co-founder Nikolai Durov to issues the warning ``Telegram channels are increasingly becoming a source of unverified information related to the Ukrainian events. We do not have the physical ability to check all the publications of the channels for reliability. I urge users from Russia and Ukraine to doubt any data that is distributed in Telegram at this time. We do not want Telegram to be used as a tool to exacerbate conflicts and incite ethnic hatred. In the event of an escalation of the situation, we will consider the possibility of partially or completely limiting the work of Telegram channels in the countries involved for the duration of the conflict.'' Shortly after posting this message on Telegram, however, Durov, walked back his threat to limit access to channels.} Given its popularity and impact on the current disinformation landscape, we began our analysis with this channel and recursively scraped all channels discovered through the forwarded messages feature. The open-source Python library Telethon\footnote{\url{https://pypi.org/project/Telethon}} was used to scrape these channels.

\begin{figure*}
    \centering
    \begin{tabular}{c}
        \includegraphics[width=0.95\linewidth]{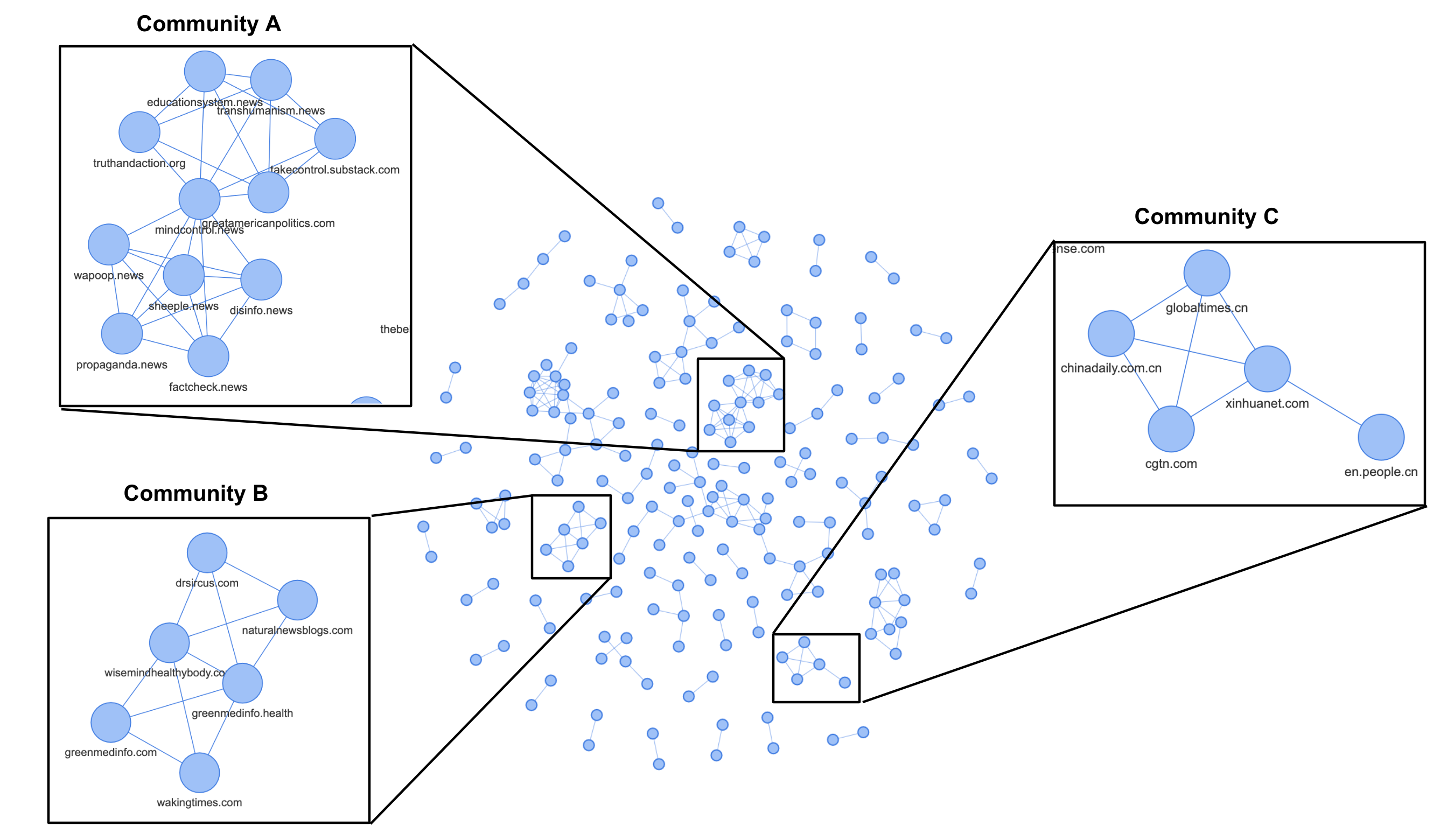}
    \end{tabular}
    \caption{The mutual Twitter-sharing graph of disinfo domains. Each node is a disinfo domain and has an undirected edge if the Jaccard index of users sharing the respective domains is at least $3\%$ (nodes without connections are excluded). The insets are the results of a community detection algorithm highlighting domains having a disproportionately high overlap in the users sharing them.}
    \label{fig:twitter}
\end{figure*}

As with the domain-level graph in Figure~\ref{fig:in_out_degree}, a graph was constructed in which each node is a telegram channel, and a directed edge connects node {\em A} to node {\em B} if channel {\em A} forwarded a message to channel {\em B}. This recursive scraping of $1,802$ channels yielded a total of $11,127$  channels. From these channels, a total of $1,013,034$ messages were collected from which $147,584$ contained a shared domain URL, $834$ of which were contained in our disinfo data set.

To understand who shares misinformation on Telegram and their relationship to each other, the graph was partitioned -- using the Louvain method~\cite{blondel2008fast} into communities based on channel-to-channel connectivity. This process yielded two broad communities shown in Figure~\ref{fig:telegram-communities}: (1) Russian, constituting $59.8\%$ of all channels; and (2) the US Far Right, constituting $40.2\%$ of all channels. The Russian community are further composed of two communities consisting of Russian media channels and Russian influencers. The US Far-Right community contains notable QAnon followers such as Sidney Powell (best known as an avid conspiracy theorist and for her part in trying to overturn the 2020 US presidential election) and QAnon enthusiast Rep. Marjorie Taylor Green.

We find that the channel {\tt SidneyPowell} is the most prolific sharer of content found in our disinfo data set. A total of $61$ new domains were scraped from $1,000$ recent posts from this Telegram channel. Of these $61$ domains, $21$ domains not in our original data set are classified by our amalgamated classifier (Section~\ref{subsec:amalgamated}) as disinfo, and manually verified by us to be clear disinfo domains. These domains include: {\tt independentside.com}, {\tt dailynewsbreak.org}, and {\tt freeworldnews.tv} (Media Bias/Fact rates this cite as ``Extreme Right Biased and a Tin-Foil Hat Conspiracy and Quackery level Pseudoscience source based on the routine publication of disinformation and outright fake news.''  One site, {\tt madison.com}, was incorrectly flagged as disinfo.

After the {\tt SidneyPowell} channel, the next four most prolific disinfo channels are {\tt worlddoctorsalliance}, {\tt RealGenFlynn}, \linebreak {\tt RealEMichaelJones}, and {\tt DrLynnFynn}.

By searching public Telegram channels for identified disinfo domains, and then performing a connectivity analysis similar to our earlier hyperlink analysis, we can identify prolific pushers of disinformation (and their relationship to each other) not just at the domain level, but also at the Telegram channel level. This analysis also allows us to grow our disinfo domain list by collating and classifying previously unseen domains from these problematic channels. Private channels, of course, would not be amenable to this type of analysis.

\subsection{Twitter}
\label{subsec:twitter}

We next investigate the ability to generalize our methods to identify disinfo domains by tracking the hyperlinks shared by certain social-media users. Because of the relative ease of access, we focus on Twitter’s publicly available user data using two of Twitter's APIs (using Python's Tweepy Library\footnote{\url{https://docs.tweepy.org/en/stable/}}). 

First, the Search Tweets API
\footnote{\url{https://developer.twitter.com/en/docs/twitter-api/v1/tweets/search/guides/standard-operators}} allows for curating a set of recent tweets (up to $9$ days old), which can be filtered by the keywords, hashtags or the URLs they contain. Starting with $2500$ NewsGuard disinfo domains (this is an updated version of the database as compared to that used in the main analysis), we collected up to a maximum of $1000$ tweets per domain, surfacing which Twitter users were sharing these disinfo domains. 

Second, the Get Tweet Timelines API \footnote{\url{https://developer.twitter.com/en/docs/twitter-api/v1/tweets/timelines/api-reference/get-statuses-user_timeline}} allows for curating up to $3200$ most recent tweets posted by a queried Twitter user. Using this API, we collect the recent domains shared by users surfaced in the first step. The data returned by both APIs contains additional user attributes such as geo-location and tweet attributes such as timestamp and replied-to, that could be leveraged in the future.

We found $569$ of the $2500$ disinfo domains had been shared at least once on Twitter. The top five most shared domains and [number of unique tweeters] are:
\begin{enumerate}
    \item {\tt organicconsumers.org} [984] 
    \item {\tt aflds.org} [981]
    \item {\tt newspunch.com} [979]
    \item {\tt mintpressnews.com} [941]
    \item {\tt icandecide.org} [934]
\end{enumerate}

An undirected domain-level graph was constructed (Figure~\ref{fig:twitter}) in which each node is a  disinfo domain, and an undirected edge connects node $A$ to node $B$ if atleast $3\%$ of the users sharing either domain also share both domains (i.e.,~ the Jaccard similarity $\ge$  $3\%$). The resulting graph contains $194$ nodes and $211$ edges. A community detection algorithm~\cite{cordasco2010community} was employed with the aim of finding if clusters of domains shared by the same users would reveal a common theme in the content being shared.

This community analysis revealed some interesting clusters. Eight of the total $11$ domains from Community {\em A} are part of the Natural News Network, included in the NewsGuard disinfo dataset, and whose articles have failed fact checks by fact-checking organizations and are described by Wikipedia as ``fake news website'' spreading pseudoscience and disinformation. Community {\em B} is dominated by health pseudoscience spreaders, of which {\tt naturalnewsblogs.com} is also part of the Natural News Network. Community {\em C} contains some Chinese state-controlled media, which have been accused of spreading Russian disinformation in the Ukraine war.

To find what other domains are shared by prolific disinformation spreaders, we scraped the Twitter activity of the top-$1000$ users sharing the most number of unique disinfo domains. Outside of {\tt twitter.com}, the top-five most shared domains and [number of unique tweeters] are:
\begin{enumerate}
    \item {\tt youtube.com} [106] 
    \item {\tt foxnews.com} [74]
    \item {\tt breitbart.com} [69]
    \item {\tt rumble.com} [68]
    \item {\tt nypost.com} [64]
\end{enumerate}
Along with online video platforms like {\tt youtube.com} and {\tt rumble.com}, hyperlinks to particularly partisan news websites also feature prominently in their tweets.

\section{Discussion}

We should begin our discussion with an acknowledgement that the topic of disinformation can and has become highly political. This can make research into and proposals to mitigate disinformation highly volatile. The political right may accuse us of bias because right-leaning  organizations may be more implicated in spreading election-related misinformation, and the political left may accuse us of bias because left-leaning organizations may be more implicated in spreading others forms of misinformation. We have, throughout, tried to remain politically neutral. From the very beginning of our research, we relied on established and independent organizations to seed our analysis in the form of the initial disinfo domains supplied by NewsGuard and GDI. Throughout our research and analysis, we then allowed the data to drive the results and conclusions.

With some $500$ hours of video uploaded to YouTube every minute, and over a billion posts to Facebook each day, the massive scale of social media makes tackling misinformation (the unintentional spreading of false or misleading information) an enormous challenge. Tackling disinformation (the intentional spreading of false or misleading information), however, may be more tractable. 

Previous studies, for example, have found that a relatively small number of users are responsible for the majority of COVID and climate-change misinformation~\cite{salam2022majority,paul2021climate}. We propose that tackling disinformation at the domain-level similarly narrows the scope of problematic content to a more manageable number. In particular, we propose that search engines and recommendation algorithms can simply demote content that contains links to previously documented (and periodically reviewed) domains. Similarly, Telegram, Twitter, TikTok, etc. channels that predominantly share problematic content can be demoted, possibly demonetized, and only in extreme cases deplatformed.

We understand and appreciate the need to balance an open and free internet, where ideas can be debated, with the need to protect individuals, societies, and democracies. Social media, however, cannot hide behind the facade they are creating a neutral marketplace of ideas where good and bad ideas compete equally. They do not. It is well established that social media's recommendation algorithms favor the divisive, outrageous, and conspiratorial because it increases engagement and profit~\cite{crockett2017moral}. As a result, Brandies' concept~\cite{curtis1996free} that the best remedy for falsehoods is more speech, not less, simply doesn't apply in the era of algorithmic curation and amplification. By demoting disinformation peddlers, we can create a more fair and balanced marketplace of ideas where Brandeis' principle can, in fact, flourish.

As with any inherently adversarial relationship, all approaches to addressing disinformation -- including ours -- will have to adapt to new and emerging threats. In our case, disinformation peddlers may add decoy hyperlinks to external trustworthy domains to escape being classified based on their hyperlinks to other disinformation domains. This, in turn, will require techniques to root out such decoy links. And so on, and on, and on. While such a cat and mouse game can be frustrating, the end game will be that it will become increasingly more difficult and time consuming to intentionally create and spread false and misleading information, with the eventual goal of discouraging most, leaving us to contend with the die-hard adversary. While this is not a complete success, it will mitigate the risk of disinformation and, hopefully, return some civility and trust to our online ecosystems.


\section*{Data and Code}

All of the code associated with this research is available in an anonymous repository: \url{https://anonymous.4open.science/r/disinfo_detection-96BA/README.md}

The data, consisting of the disinfo domains, are licensed from NewsGuard and GDI and we therefore cannot make this data publicly available per the licensing agreement terms.

\bibliographystyle{ACM-Reference-Format}
\bibliography{main}

\end{document}